\documentstyle[prl,aps,psfig]{revtex}

\begin{document}

\twocolumn[\hsize\textwidth\columnwidth\hsize\csname
@twocolumnfalse\endcsname

\draft
\preprint{Submitted to Phys. Rev. Lett.}
\title{Extremely High Energy Neutrinos, Neutrino Hot Dark Matter,\\
and the Highest Energy Cosmic Rays}
\author{Shigeru Yoshida
\cite{byline}
}
\address{Institute for Cosmic Ray Research, University of Tokyo,
Tanashi, Tokyo 188-8502, Japan}
\author{G\"unter Sigl and Sangjin Lee}
\address{Department of Physics \& Astrophysics, 
Enrico Fermi Institute, The University of Chicago, Chicago, IL 60637-1433}
\date{Submitted to Phys.Rev.Lett.}
%
\maketitle
%
\begin{abstract}
Extremely high energy ($\sim10^{22}$ eV) cosmic neutrino beams
initiate high energy particle cascades in the
background of relic neutrinos from the Big Bang.
We perform numerical calculations to show that such cascades
could contribute more than 10\% to the observed cosmic ray flux
above $10^{19}\,$eV if neutrinos have $\sim$ eV masses.
The required intensity of primary neutrinos 
could be consistent with astrophysical models for
their production {\it if} the maximum neutrino energy
reaches to $\sim 10^{22}$ eV {\it and} the massive neutrino
dark matter is locally clustered. Future observations of
ultra high energy cosmic rays will lead to an
indirect but practical search for neutrino dark matter. 

\end{abstract}
\vskip1pc]


It has been claimed that pure cold dark matter (CDM)
leads to a larger baryon fraction ($\Omega_{b}$) than
predicted by big bang nucleosynthesis (BBN) if the observed
hot X-ray-emitting gas represents a fair sample of the universe
\cite{strickland97}. An admixture of hot dark matter
(HDM) with CDM shifts the estimates of the baryon fraction closer
to that by BBN. In addition, this mixed cold + hot dark matter
model (CHDM) has been shown to agree well with the cosmic
microwave background (CMB) spectrum measured by COBE,
and galaxy group properties such as the number density of clusters
\cite{primack95}. Neutrinos are the best candidate for HDM
and a total neutrino mass of 5 eV, or 
$m_{\nu_\mu}\sim m_{\nu_\tau}\sim 2.4$ eV ($\Omega_{\nu}\simeq 0.2$)
may be a solution consistent with all available
observations.

If HDM consists of cosmological background neutrinos
(CBN) with $\sim$ eV mass {\it and} there exist
cosmic neutrino beams reaching to $\sim 10^{22}$ eV,
the interactions of these extremely high energy (EHE) cosmic
neutrinos with the CBN 
during their propagation can become significant
\cite{roulet93} due to the  enhanced interaction
probability at the Z boson resonance. 
The resulting neutrino cascade
causes modifications such as a bump and a dip in the 
EHE neutrino spectrum at Earth \cite{yoshida97}.
The cascade contains several hadronic decay channels
that produce mostly pions which in turn reproduce
neutrinos through their decay \cite{yoshida94}, 
but also $\gamma$-rays and some nucleons.
Hence this mechanism has been proposed
\cite{weiler97,fargion97} as an explanation of the highest energy
cosmic rays (HECRs) whose flux above $\sim5\times10^{19}$ eV
is severely attenuated by photopion
production on the cosmic microwave background (CMB) \cite{yoshida98}
in case of nucleon primaries, 
forming the Greisen-Zatsepin-Kuzmin (GZK) cutoff \cite{greisen66},

The evolution of cascades initiated by EHE cosmic neutrinos
is determined by very complex chains of interactions: 
The neutrinos undergo $\nu\nu$ reactions which involve the
exchange of W and Z bosons and hadronization of their
strongly interacting decay products.
The produced photons, electrons and protons collide with
the CMB, the infrared and optical background (IR/O)
and the universal radio background (URB) \cite{sigl97,protheroe96} 
initiating electromagnetic cascades. Electrons are also
subject to synchrotron cooling in extragalactic magnetic
fields (EGMF). The final particle fluxes after propagation
depend on all these interactions and solving the relevant
transport equations is inevitable for an accurate evaluation
of the consequences of this scenario. In this letter we
present numerical calculations of the ``primary'' EHE
neutrinos and the ``secondary'' $\gamma$-rays and protons
that may contribute a sizable fraction of the observed
HECRs above $\simeq10^{19}$ eV, under different assumptions concerning 
neutrino mass and local density enhancement of the HDM.
Several observable signatures to confirm or rule out this scenario
are discussed followed by a summary.

{\it Cascading Calculation.}--
Our numerical calculations combine
simulation codes for neutrino cascades \cite{yoshida97,yoshida94},
and for electromagnetic cascades and nucleon propagation
\cite{sigl97,lee96}. Detailed accounts
of these codes are provided in Refs. \cite{yoshida97,lee96}.
The following processes are included: Inelastic and elastic
$\nu\nu$ collisions involving an exchange of either a W or a Z
boson on the CBN; the subsequent decays of produced $\pi$'s, $\mu$'s, and
$\tau$'s, hadronization of quark jets, all of which eventually 
feed into the electromagnetic, neutrino, and nucleon channels; 
$\gamma\gamma\to e^+e^-$ on the CMB, the IR/O and the URB; inverse Compton
scattering on the same backgrounds; triplet pair production and
double pair production on the CMB; synchrotron cooling in
the EGMF; the nucleon interactions
on the CMB (pair production and photopion production),
and neutron decay; redshifting and evolution of
the black body temperature due to expansion of the universe.
For the IR/O we used recent data \cite{fixsen97}, and for the
(unmeasured) URB we used the highest prediction of
Ref. \cite{protheroe96-2} yielding conservatively low
$\gamma$-ray fluxes around $10^{20}$ eV for which the URB
is the most important target for pair production.
We neglect interactions of EHE neutrinos with the CMB photons
which are of comparable importance to those 
with the CBN only for neutrino energies above the Z
resonance \cite{seckel98}.

The hadronic decay of Z bosons resonantly produced by neutrinos
of energy
\begin{equation}
E_{\rm res}=M_z^2/2m_{\nu}= 
4\times10^{21} \left({m_{\nu}\over 1eV}\right)^{-1}eV, \label{resonance}
\end{equation}
with the CBN is the most important neutrino process for production
of $\gamma$-rays and nucleons whose spectra are determined by
the hadron fragmentation function. At the
energy range around the Z pole, this has been measured
accurately by the LEP at CERN. We implemented empirical
functions using the MLLA approximation \cite{khoze97}
in our code, which have been fitted by
measurements of the inclusive production rates
of $\pi^\pm$ and $p{\bar p}$ with the OPAL detector \cite{akers94}. 
This constitutes the major revision of the original codes
in Refs. \cite{yoshida97,yoshida94}.

The dominant contribution to secondary particle fluxes
from resonant Z production can be estimated analytically,
for example for the produced $\gamma$-ray spectrum,
\begin{equation}
{dN_{\gamma}\over dE_{\gamma}dL}\sim 
\sqrt{2}\pi {\Gamma_Z\over M_Z}{1\over \lambda_Z}
{dN_{\nu}\over dE_{\nu}}\vert_{E_{\nu}=E_{res}}
{dn_h\over dx}\vert_{x=E_\gamma/E_{res}}, \label{rate}
\end{equation}
where $\Gamma_Z\simeq0.03M_Z$ is the decay width of the Z boson
whose mass is $M_Z$ , $dn_h/dx$ is
the $\pi^0$ fragmentation spectrum, and $\lambda_Z\simeq$ 38 Gpc
is the mean free path of neutrinos at $E_{\rm res}$ given by
Eq.~(\ref{resonance}). The number of photons above
$3\times 10^{19}/(m_\nu/\,{\rm eV})$ eV is then
$0.012\,E_{\rm res}(dN_\nu/dE_\nu)(E_{\rm res})
\,{\rm Mpc}^{-1}$, compared to 0.03 from our full numerical
calculation which includes
contributions from all channels and uses
the more accurate MLLA formula for the fragmentation function.

{\it The particle fluxes}.--
For a general discussion
we consider a homogeneous distribution of sources
radiating EHE neutrinos with a constant differential spectrum
$\propto E^{-q}$ and a luminosity per comoving volume that scales as
$(1+z)^m$ between $z=z_{\rm min}$ and $z=z_{\rm max}$ \cite{yoshida97},
with $m$ characterizing source evolution.
Because of the small neutrino absorption probability, the
results are essentially independent of $z_{\rm min}\lesssim1$.
We assume a flat universe with a Hubble constant
of $H_0= 65$ km sec$^{-1}$ Mpc$^{-1}$ which is consistent
with the CHDM picture of the universe \cite{strickland97,primack95}.
We use $q=1$, a typical spectral index expected for neutrinos
produced from photopion production by accelerated protons
\cite{stecker96,halzen97,waxman98}. Results are, however, rather
insensitive to $q$ for $q\lesssim2$. The ratio of
emitted $\nu_{\mu}$ and $\nu_e$ fluxes is assumed to
be $\simeq1.86$, as expected from charged pion decay.
Furthermore, for the moment we assume that the source
luminosity in $\gamma$-rays and nucleons is
negligible compared to the neutrino luminosity.

HDM is usually expected to cluster locally and the Fermi
distribution with a velocity dispersion $v$ yields
an overdensity
$f_\nu\lesssim v^3 m_\nu^3/(2\pi)^{3/2}/\bar{n_\nu}\simeq
330\,(v/500\,{\rm km}\,{\rm sec}^{-1})^3\,(m_\nu/{\rm eV})^3$
over the uniform CBN \cite{peebles}. 
If clustering occurs on a scale $l_\nu$ smaller
than the typical attenuation length $l_{\rm att}\sim$ few Mpc
of nucleons and $\gamma$-rays around $10^{20}$ eV, the ratio
of their fluxes produced
on that scale to the ones produced on the uniform background
is $\simeq f_\nu l_\nu/l_{\rm att}$. Therefore, while
clustering in the galactic halo or in a nearby galaxy
cluster is unlikely to contribute to the HECR flux
\cite{waxman98a}, neutrinos clustering in the local
Supercluster may have $f_\nu\sim100$, $l_\nu\sim$ few
Mpc.

\begin{figure}[tb]
\centerline{\psfig{file=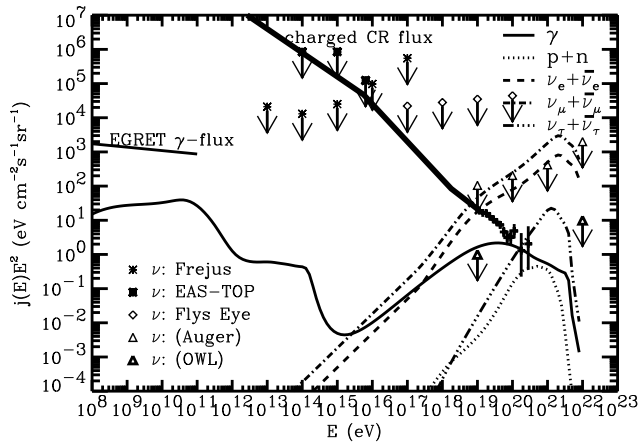,height=6cm}}
\medskip
\caption[...]{Energy spectra of nucleons, $\gamma$-rays and neutrinos
for the scenario described in the text. Error bars are the combined data 
from the Fly's Eye \cite{bird} and the AGASA \cite{takeda}
experiments above $10^{19}$ eV. Also shown are piecewise power
law fits to  the observed charged CR flux below $10^{19}$ eV, the
EGRET measurement of the diffuse $\gamma$-ray flux between
30 MeV and 100 GeV, and experimental neutrino flux limits from
Frejus \cite{rhode96} and Fly's Eye \cite{baltrusaitis}, as well
as projected neutrino sensitivities of the future Pierre Auger
\cite{capelle98} and NASA's OWL \cite{ormes98} projects.}
\label{fig:fig1}
\end{figure}

In Fig. \ref{fig:fig1} we show
the calculated spectra for the following typical case:
$m=3$, $z_{\rm min}=0$, $z_{\rm max}=3$, an EGMF
$\lesssim10^{-12}$ Gauss, $m_{\nu_e}=m_{\nu_\mu}=m_{\nu_\tau}=1$ eV, 
$f_\nu\simeq300$, and $l_\nu=5$ Mpc.
It can be seen clearly that
the predicted fluxes are consistent with the measurement of the
diffuse $\gamma$-ray flux by EGRET \cite{sreekumar97}
and with upper limits on
neutrino fluxes by Frejus \cite{rhode96} and Fly's Eye \cite{baltrusaitis}.
Typically, the energy content in the produced low energy
cascade $\gamma$-rays is a few percent of the neutrino energy
which agrees with a rough analytical estimate giving
$\sim10/(H_0\lambda_Z)\Gamma_Z/M_Z$.
By scaling the cosmologically produced low
energy $\gamma$-ray flux in Fig. \ref{fig:fig1} with
$l_{\rm att}/(f_\nu l_\nu)$, the
EGRET constraint on diffuse $\gamma$-rays requires
$f_\nu\gtrsim20\,(l_\nu/5\,{\rm Mpc})^{-1}$.

The EHE part of the secondary $\gamma$-rays and protons
possibly constitute a hard component of the observed HECRs without
a GZK cutoff. The energy content in this ``visible'' HECR
component is about $(\Gamma_Z/M_Z)(f_\nu l_\nu/\lambda_Z)
E_{\rm res}^2(dN_\nu/dE_\nu)(E_{\rm res})$, again consistent with
the fluxes shown in Fig. \ref{fig:fig1}.
The collisions of the EHE cosmic neutrinos
with the HDM can be responsible for $\sim10\%$ of the observed
cosmic rays above 
$10^{19}$ eV with dominant contributions of $\gamma$-rays above the
GZK cutoff. The fluxes deviate at most by 50\% for
$m_{\nu_e}\ll m_{\nu_\mu}$.

\begin{figure}[tb]
\centerline{\psfig{file=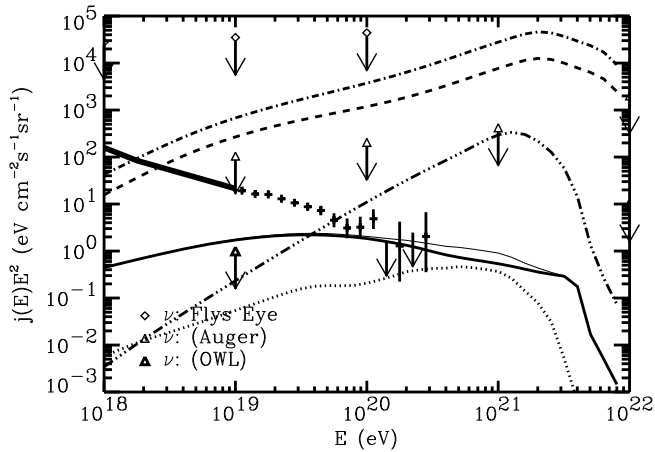,height=6cm}}
\medskip
\caption[...]{Same as in Fig. \ref{fig:fig1} but for the case of
an overdensity of 20 over 5 Mpc, showing fluxes above $10^{18}$ eV.
Line key as in Fig. \ref{fig:fig1} with the ``visible'' sum
of $\gamma$-ray and nucleon fluxes shown as thin solid line
in addition, as well as upper limits in the bins where
no HECR were seen below the highest energy event. This case
sets an upper bound for the intensity of primary EHE neutrino beams
allowed by the EGRET diffuse $\gamma$-ray limit, assuming only
secondaries of neutrino interactions contribute to the EGRET
flux.}
\label{fig:fig2}
\end{figure}

Fig. \ref{fig:fig2} shows the high energy part of the
resultant spectra above $10^{18}$eV as in 
Fig. \ref{fig:fig1}, but for the case of the lower local enhancement
of the neutrino dark matter, $f_\nu=20$ over a scale $l_\nu=5$ Mpc,
the lowest possible $f_\nu$ allowed by the EGRET bound. As
compared to the case for stronger clustering shown in
Fig. \ref{fig:fig1}, the required EHE neutrino intensity is
10 times larger.

\begin{figure}[tb]
\centerline{\psfig{file=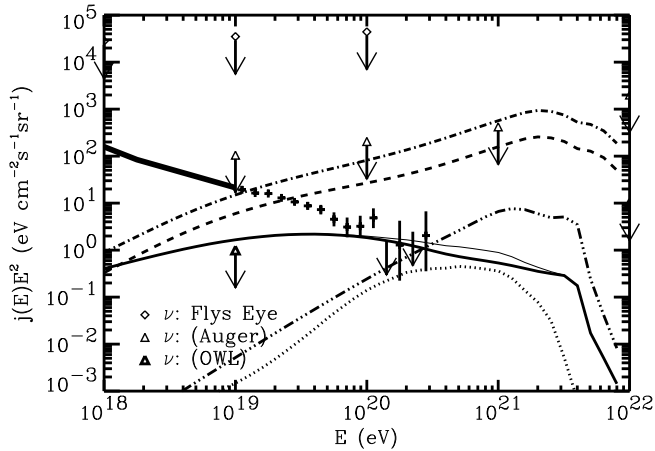,height=6cm}}
\medskip
\caption[...]{Same as in Fig. \ref{fig:fig2} but for the
optimistic case of an overdensity of $10^3$ over 5 Mpc
that would be required if $L_\gamma\simeq13L_\nu/3$.}
\label{fig:fig3}
\end{figure}

In general, models based on photopion production predict an
integrated photon source luminosity $L_\gamma$ that is comparable to
the total neutrino luminosity, $L_\gamma\simeq\frac{13}{3}L_\nu$
\cite{halzen97,mannheim95}. In this case, the EGRET constraint
translates into the more stringent requirement $f_\nu\gtrsim
10^3(l_\nu/5\,{\rm Mpc})^{-1}$, as can be seen by applying the
above mentioned scaling to the integrated neutrino luminosity
from Fig. \ref{fig:fig1}. 
Fig. \ref{fig:fig3} shows
the fluxes for this optimistic case of strong clustering.
This bound on $f_\nu$ can be relaxed if most of $L_\gamma$
does not appear at EGRET energies, but is dominantly
released in the TeV range. This could be a detectable
signature from individual point sources \cite{aharonian94}, 
in addition to the secondary $\gamma$-rays from neutrino interactions 
appearing at EGRET energies.
Furthermore, the scenario discussed here requires sources that
are optically thick for accelerated protons with respect to
photopion production because otherwise the observable
proton flux below the GZK cutoff would be comparable
to the neutrino flux \cite{waxman98}.


{\it Discussion.}--
The EHE neutrino scenario we explored here is quite solid
in terms of the particle physics because the interactions with
the cosmological backgrounds occur in the well measured
LEP energy range. No physics beyond the Standard
Model is involved except neutrino mass.
The major uncertainty arises in the question whether
any astronomical objects are capable of producing neutrinos
with energies of $\sim E_{\rm res}$. In the conventional
models invoking the decay of photoproduced pions,
primary protons must be accelerated
to $\sim20E_{\rm res}\sim 10^{23}/(m_\nu/{\rm eV})$ eV 
in order to generate neutrinos of energy $E_{\rm res}$.
Furthermore, the sources would need a dense photon target
to supply high neutrino luminosity and to absorb protons 
and $\gamma$-rays. Thus, a new model for the neutrino beam
sources may be necessary \cite{takahashi98}.

Interestingly,
the energy generation rate of $\sim E_{\rm res}$ neutrinos
for the scenario shown in Fig. \ref{fig:fig1},
$1.8\times 10^{45}$ erg Mpc$^{-3}$ yr$^{-1}$ divided by 
the the rate of cosmological Gamma Ray Bursts (GRBs),
$3\times 10^{-8}$ Mpc$^{-3}$ yr$^{-1}$ \cite{mao92}, 
yields $\sim 6\times 10^{52}$ erg, and is comparable to the observed 
energy release including afterglow from a typical GRB in the BATSE range
\cite{waxman97}.

The EHE neutrino scenario has several advantages to
explain the HECR observation.
The observed relatively hard spectrum without GZK cutoff \cite{takeda}
is reasonably reconstructed in our model which is determined mainly by
the well-measured hadron fragmentation function at the Z pole
and the energy loss process in the cosmological backgrounds,
regardless of the nature of the EHE neutrino sources.
The highest energy events above the GZK cutoff can originate
from very distant powerful objects because neutrinos propagate
without significant energy loss.  For example, the AGN 3C147
at a redshift of 0.545 is a candidate for the Fly's Eye
$3\times 10^{20}$ eV event \cite{elbert95}.
For the same reason, it is natural that we found no 
{\it nearby} powerful astronomical objects in directions of
the possible event clusters observed by AGASA \cite{hayashida96}.
Because the EHE neutrino beams can be responsible for a
sizable fraction of cosmic rays above $10^{19}$ eV,
this scenario can explain
the observational fact that the intensity of the
events observed above the GZK cutoff is consistent with
the extrapolation of the flux from lower energies.

Among the observable signatures of the neutrino scenario are
the primary EHE neutrinos whose flux should be detectable,
as projected sensitivities of future experiments
such as the Pierre Auger \cite{capelle98} and NASA's OWL \cite{ormes98}
projects suggest.
Correlation of the arrival direction of the EHE neutrino and
the secondary HECR showers may also be observable.
As opposed to conventional models with nucleon primaries,
our model predicts that some of the observed HECRs should originate
in sources at cosmological distances.
At energies beyond the GZK cutoff, the correlation
of arrival directions of HECR showers with sources at distances
$\gg l_{\rm att}$ should be easy to detect since the background
from a conventional nucleon component should be suppressed
due to the GZK effect, whereas the component proposed here
continues as a relatively flat spectrum. Finally, this scenario
predicts a $\gamma$-ray domination above $10^{20}$ eV, and next
generation experiments should settle the question whether
observed HECR are consistent with $\gamma$-ray primaries
\cite{elbert95}.

In summary, we have seen that collisions of EHE cosmic neutrino
beams with $\sim$ eV mass neutrino dark matter would explain the observed
HECR energy spectrum, regardless of
the nature of the neutrino sources if the maximum neutrino energy
reaches to the Z boson pole region and the dark matter
is clustered on the Supercluster scale by amounts consistent with 
expectations. Although EHE neutrino sources require
very high efficiency of converting the energy to neutrino flux
which may require a new production mechanism of neutrinos,
the necessary neutrino intensity can be consistent with
observed diffuse $\gamma$-ray fluxes and the GRB energy release rate. 
The EHE neutrino scenario is a way of producing a relatively
flat component of nucleons and $\gamma$-rays that provides
a significant fraction of the HECR flux above $10^{19}$ eV,
dominating above the GZK cutoff, without invoking
physics beyond the Standard Model except neutrino mass. 
Future observations of HECRs lead to indirect search for signatures 
of neutrino dark matter.

We thank Eli Waxman and Makoto Sasaki for helpful discussions and advice
and Tom Weiler and Pijush Bhattacharjee for encouragement.
Paolo Coppi is acknowledged for earlier collaboration
on electromagnetic cascades.
At the University of Chicago this research was
supported by the DoE, NSF, and NASA.


%
%

%
%


\begin{references}

\bibitem[*]{byline}  Electronic Address: syoshida@icrr.u-tokyo.ac.jp

\bibitem{strickland97}
R. W. Strickland and D. N. Schramm, Astrophys. J.  {\bf 481}, 571 (1997).

\bibitem{primack95} 
J. R. Primack {\it et al.}, Phys. Rev. Lett. {\bf 74}, 2160 (1995).

\bibitem{roulet93} E. Roulet, Phys. Rev. D {\bf 47}, 5247 (1993).

\bibitem{yoshida97} 
S. Yoshida {\it et al.}, Astrophys. J. {\bf 479}, 547 (1997).

\bibitem{yoshida94} S. Yoshida, Astropart. Phys. {\bf 2}, 187 (1994).

\bibitem{weiler97} T. J. Weiler, hep-ph/9710431, to be published
in Astropart. Phys.

\bibitem{fargion97} D. Fargion, B. Mele, and A. Salis, astro-ph/9710029.

\bibitem{yoshida98}
For reviews, see S. Yoshida and H. Dai, J. Phys. G {\bf 24}, 905 (1998). 

\bibitem{greisen66}
K. Greisen, Phys. Rev. Lett. {\bf 16}, 748 (1966); 
G. T. Zatsepin and V. A. Kuzmin, JETP Lett. {\bf 4}, 178 (1966).

\bibitem{sigl97} G. Sigl {\it et al.}, Phys. Lett. B {\bf 392}, 129 (1997).

\bibitem{protheroe96}
R. J. Protheroe and T. Stanev, Phys. Rev. Lett. {\bf 77}, 3708 (1996);
erratum Phys. Rev. Lett. {\bf 78}, 2420 (1997).

\bibitem{lee96}
S. Lee, Phys. Rev. D {\bf 58}, 043004 (1998).

\bibitem{fixsen97} D. J. Fixsen {\it et al.}, Astrophys. J. {\bf 490},
482 (1997).

\bibitem{protheroe96-2} R. J. Protheroe and P. L. Biermann,
Astropart. Phys. {\bf 6}, 45 (1996).

\bibitem{seckel98}
D. Seckel, Phys. Rev. Lett. {\bf 80}, 900 (1998).

\bibitem{khoze97}
V. A. Khoze and W. Ochs, Int. J. Mod. Phys. A {\bf 12},
2949 (1997). 

\bibitem{akers94}
R. Akers {\it et al}., Z. Phys. {\bf C63}, 181 (1994).

\bibitem{stecker96} F. W. Stecker and M. H. Salamon, 
Space. Sc. Rev. {\bf 75}, 341 (1996).

\bibitem{halzen97}
F. Halzen and E. Zas, Astrophys. J.  {\bf 488}, 669 (1997).

\bibitem{waxman98}
E. Waxman and J. Bahcall, hep-ph/9807282, submitted to Phys. Rev. D.

\bibitem{peebles} see, e.g., P. J. E. Peebles, {\it Principles of Physical
Cosmology}, Princeton University Press, New Jersey, 1993.

\bibitem{waxman98a} E. Waxman, astro-ph/9804023, submitted to
Astropart. Phys.

\bibitem{sreekumar97} P. Sreekumar {\it et al.},
Astrophys. J. {\bf 494}, 523 (1998).

\bibitem{rhode96} W. Rhode {\it et al.}, Astropart. Phys. {\bf 4}, 217 (1996).

\bibitem{baltrusaitis} R. M. Baltrusaitis {\it et al.},
Phys. Rev. D {\bf 31}, 2192 (1985).

\bibitem{bird} D. J. Bird {\it et al.}, 
Phys. Rev. Lett. {\bf 71}, 3401 (1993).

\bibitem{takeda} M. Takeda {\it et al.} 
Phys. Rev. Lett. {\bf 81}, 1163 (1998).

\bibitem{capelle98} K. S. Capelle, J. W. Cronin, G. Parente,
and E. Zas, Astropart. Phys. {\bf 8}, 321 (1998).

\bibitem{ormes98} J. F. Ormes {\it et al.}, in Proc. 25th
International Cosmic Ray Conference
(Durban, 1997), eds.: M. S. Potgieter {\it et al.}

\bibitem{mannheim95} 
K. Mannheim, Astropart. Phys. {\bf 3}, 295 (1995).

\bibitem{aharonian94}
F.A. Aharonian, P.S.Coppi, and H.J.V\"olk, Astrophys.J. {\bf 423}, L5 (1994).

\bibitem{takahashi98}
One possible mechanism to produce neutrinos efficiently
without EHE proton radiation
is the collision of protons directly accelerated by 
Wakefield and Snowplow plasma
at initial GRB fireball with thick $\gamma$-rays. 
See Y. Takahashi in Proceedings of
Workshop on Observing Giant Air Showers from Space, edited by 
J. F. Krizmanic {\it et al.} (The American Institute of Physics, 1998),
p. 469.

\bibitem{mao92}
S. Mao and B. Paczy\'nski, Astrophys. J. {\bf 388}, L45 (1992);
E. Cohen and T. Piran, Astrophys. J. {\bf 444}, L25 (1995).

\bibitem{waxman97}
E. Waxman and J. Bahcall, Phys. Rev. Lett. {\bf 78}, 2292 (1997) 
and references therein;
M. Vietri, Phys. Rev. Lett. {\bf 80}, 3690 (1998).

\bibitem{elbert95}
J. W. Elbert and P. Sommers, Astrophys. J.  {\bf 441}, 151 (1995);
F. Halzen {\it et al}., Astropart. Phys. {\bf 3}, 151 (1995).

\bibitem{hayashida96}
N. Hayashida {\it et al}., Phys. Rev. Lett. {\bf 77}, 1000 (1996).

\end{references}
\end{document}